\documentclass[journal]{IEEEtran}
\usepackage{url}
\usepackage{hyperref}
\usepackage{graphicx}
\usepackage{amssymb}
\usepackage{amsmath}
\usepackage{setspace}
\usepackage{graphics} 
\usepackage{graphicx}
\usepackage{epsfig}
\usepackage{caption}
\usepackage[sorting=none]{biblatex}
\bibliography{MasterBiblio}

\graphicspath{{./Figures/}}

\newcommand{\be}{\begin{equation}} 
\newcommand{\ee}{\end{equation}}   

\newcommand{\bes}{\begin{equation*}} 
\newcommand{\ees}{\end{equation*}}   

\newcommand{\beas}{\begin{eqnarray*}} 
\newcommand{\eeas}{\end{eqnarray*}}   

\newcommand{\bea}{\begin{eqnarray}} 
\newcommand{\eea}{\end{eqnarray}}   

\newcommand{\ba}{\begin{array}}
\newcommand{\ea}{\end{array}}

\newcommand{\bbm}{\begin{bmatrix}}
\newcommand{\ebm}{\end{bmatrix}}

\def\ni{\noindent}

\def\k{\mathrm{k}}

\def\d1{ \left( \frac{\beta}{\gamma} \right)'}

\def\nn{\nonumber}
\def\eqd{\stackrel{\textsf{d}}{=}}

\def\R{\mathbb{R}}

\def\a{\alpha}

\def\ds{\displaystyle}

\def\Del{\Delta}

\def\R{\mathbb{R}}

{\begin{list}{}
         {\setlength{\leftmargin}{#1}}
         \item[]
}
{\end{list}}

\newcommand{\bsub}{\begin{subequations}}
\newcommand{\esub}{\end{subequations}}

\def\wbg4{\bar{w}_{\gamma_4}}
\def\wbq4{\bar{w}_{q_4}}

\def\yr{y_{R_{eq}}}
\def\yrd{\dot{y}_{R_{eq}}}
\def\yg{y_{G_{eq}}}

\def\k2{\bar{k}_2}

\def\v1h{\hat{v}_1}

\def\c2L{\cos^2(\Lambda)}

\begin{document}
%
\title{Nonlinear Equivalent Resistance-based \\ Maximum Power Point Tracking (MPPT)}

\author{\IEEEauthorblockN{Chaitanya Poolla\IEEEauthorrefmark{1},
Abraham K. Ishihara\IEEEauthorrefmark{2}}\\
\IEEEauthorblockA{ECE,
Carnegie Mellon University (SV)\\
Moffett Field, CA 94035\\
Email: \IEEEauthorrefmark{1}cpoolla@alumni.cmu.edu,
\IEEEauthorrefmark{2}abe.ishihara@west.cmu.edu}}

\maketitle
\makeatletter
\def\ps@IEEEtitlepagestyle{
  \def\@oddfoot{\mycopyrightnotice}
 \def\@evenfoot{}
}
\def\mycopyrightnotice{
  {\footnotesize
  \begin{minipage}{\textwidth}
  \centering
  \copyright~2019 IEEE. Personal use of this material is permitted. Permission from IEEE must be obtained for all other uses, in any \\ current or future media, including
reprinting/republishing this material for advertising or promotional purposes, creating new \\
collective works, for resale or redistribution to servers or lists, or reuse of any copyrighted
component of this work in other works.
  \end{minipage}
  }
}
\pagenumbering{gobble}
\begin{abstract}
We present a nonlinear equivalent resistance tracking method to optimize the power output for solar arrays. Tracking an equivalent resistance results in nonlinear voltage step sizes in the gradient descent search loop. We introduce a new model for the combined solar module along with a DC-DC converter which results in a highly nonlinear dynamical system due to the inherent non-linearity of the PV cell topology and the switched DC-DC converter system. To guarantee stability over a range of possible operating regimes, we utilize a feedback linearization control approach to exponentially converge to the setpoint. Simulations are presented to illustrate the performance and robustness of the proposed technique.
\end{abstract}

\begin{IEEEkeywords} 
Solar, Photovoltaic, Maximum Power Point Tracking, MPPT, Feedback Linearization, Buck-boost converter
\end{IEEEkeywords}

\section{INTRODUCTION}
Recently, we have seen solar installations in the US more than double in every market segment \cite{solarMarketInsight2011}.  Despite this recent growth, improvements in reliability testing, and advancements in solar cell efficiency, significant questions remain regarding the actual performance of PV modules in the field.  Non-uniform changes in solar cell parameters may render modules more susceptible to hot-spot generation, especially under soiling and partial shading conditions.  In order to optimize performance, per panel Maximum Power Point Tracking (MPPT) leveraging intelligent control techniques has been shown to be a viable solution \cite{Deline_Marion_Granata_Gonzalez_2011}.

The problem of Maximum Power Point Tracking (MPPT) has been well-studied in the literature \cite{Esram:2007}. Common approaches include, Hill-Climbing, Perturb and Observe (P\&O), Incremental Conductance (IC), Fuzzy Logic (FL), Neural Networks (NN), and Ripple Correlation Control (RCC). The non-model based approaches such as Hill-Climbing and P\&O seek to estimate the sign of the gradient at the operating point on the P-V curve. Assuming uniform conditions without faults the P-V curve is known to have a single maximum. Hence, knowledge about the sign of the gradient is sufficient to determine the direction of perturbation in the voltage space. However, inappropriate choice of step sizes often lead to oscillations at or near the MPP. Incremental conductance-based approaches \cite{Hussein:1995} approximate the slope along the P-V curve as $\frac{dP}{dV}=I+V\frac{\Delta I}{\Delta V}$. Thus, measurements of instantaneous and incremental conductance are sufficient to determine the direction of perturbation. While IC is capable of tracking changing weather conditions quickly, it can result in oscillations similar to P\&O \cite{tung2006evaluation}. On the other hand, sophisticated approaches such as those based on Fuzzy Logic or Neural Networks require regular tuning for adaptability. The tracking performance of FL-based approaches depend on the choice of membership functions \cite{veerachary2003neural}. In case of NN-based approaches, the network needs to trained for a given PV array and tuned further to adapt to changing array characteristics \cite{Esram:2007}. Further, the Maximum Power Point (MPP) of the PV system depends on local weather conditions and hence accurate prediction of the environmental conditions \cite{poolla2018solar} enables better MPP tracking \cite{gao2013maximum}.

In this work, we consider a PV plant integrated with a buck-boost converter supporting a load. The schematic is shown in Figure \ref{FIG:buck_boost_test-01}. By adjusting the duty cycle of the converter, the operating point of the PV system can be driven toward the MPP. Unlike several Hill-Climbing or P\&O approaches that seek to uniformly perturb the operating voltage, we propose to perturb the equivalent resistance (or, conductance) at the operating point which results in nonlinear voltage changes. This perturbation sets up the target operating point for the inner loop tracker, which is implemented in the buck-boost converter using feedback linearization. The main contribution of this paper is the combination of an equivalent resistance tracking outer-loop based on a feedback linearization inner-loop control law that takes into account the highly nonlinear plant dynamics.

The rest of the paper is structured as follows: Section \ref{SEC:nert} provides an overview of the outer-loop iterative update along with the derivation of closed-form equations using the Lambert-W function in Section \ref{SSEC:lambertoverview}. The model of the PV module integrated with a buck-boost converter along with corresponding dynamics is described in Section \ref{SEC:solarbuckboost}. The feedback linearization controller is derived in Section \ref{SEC:FLC}. Simulation results are presented in Section \ref{SEC:results} along with conclusions in Section \ref{SEC:conclusion}.

\section{Nonlinear Equivalent Resistance Tracking}
\label{SEC:nert}
The effective or equivalent resistance seen by the solar module is given by
\be
\label{EQ:ReqDef}
R_{eq} = \frac{V}{I}
\ee
where $V$ and $I$ are the operating voltage and current of the module, respectively. Let us denote the operating point at time $t_k\in\R^{+}$ by the pair $(I_k,V_k)\in\R^2$.  Given an operating point $(I_k,V_k)$, the goal of the outer-loop MPPT algorithm is to determine a new operating point at time $t_{k+1}$ given by $(I_{k+1},V_{k+1})$ such that $I_{k+1}\cdot V_{k+1} > I_{k}\cdot V_{k}$.  If the algorithm converges, then the solar module will be operating at a local maximum of the power curve.  Under uniform conditions, there is only one global maximum and hence, gradient descent algorithms will converge to the maximum power point. 

A challenge is to determine change $(\Del I_k,\Del V_k) \eqd (I_{k+1},V_{k+1})-(I_k,V_k)$ since the power versus voltage (or current) landscape is highly nonlinear. It is common to use a constant $\Del V >0$ voltage increment or decrement depending on the slope of the power versus voltage curve, $\left.\frac{dP}{dV}\right|_{(I_k,V_k)}$. If $\left.\frac{dP}{dV}\right|_{(I_k,V_k)} >0$, $V_{k+1} = V_k+\Del V$. If $\left.\frac{dP}{dV}\right|_{(I_k,V_k)} <0$, $V_{k+1} = V_k-\Del V$.  The problem with this approach is that a constant $\Del V$ that works in one region of the I-V curve may not work in another. For example, a small $\Del V$ increment that is suitable to the left of the MPP, may be too large of an increment for operating regimes to the right of the MPP due to the large negative slope of the $I-V$ curve. 

\begin{figure}[t]
\begin{center}
\includegraphics[width=0.35\textwidth]{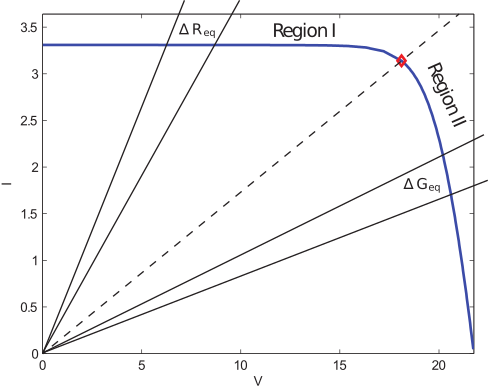} 
\end{center}
\caption{\enskip Regions I and II of I-V curve}
\label{FIG:IV}
\end{figure}

An alternative approach is to consider a constant equivalent resistance (or, conductance) change when the operating point is in region I (or, II) (Fig. \ref{FIG:IV}).  If the operating point is in region I, we increase the equivalent resistance by a constant $\Del R_{eq}$.  
\be
\label{EQ:ReqDef2}
R_{eq_{k+1}} = R_{eq_k} + \Del R_{eq}
\ee
where $\Del R_{eq}>0$ is a constant for all $t_k$.  Using (\ref{EQ:ReqDef}), (\ref{EQ:ReqDef2}) becomes
\be
\label{EQ:beta1}
\frac{V_{k+1}}{I_{k+1}} = \frac{V_{k}}{I_{k}} + \Del R_{eq} \eqd \beta_k
\ee

When the operating point of the solar module is in region II (Fig. \ref{FIG:IV}), we choose to increment by the inverse of equivalent resistance, or equivalent conductance.  That is: 
\be
\label{EQ:Geq}
\frac{I_{k+1}}{V_{k+1}} = \frac{I_{k}}{V_{k}} + \Del G_{eq} \eqd \frac{1}{\beta_k}
\ee
It again follows that (compare to (\ref{EQ:beta1}))
\be
\label{EQ:beta2}
V_{k+1} = \beta_k I_{k+1}
\ee

Given $\beta_k$ (Equations \ref{EQ:beta1} or \ref{EQ:Geq}), we may use the one-diode model of the solar array and the Lambert-W function \cite{Petrone:2007,Corless_Gonnet_Hare_Jeffrey_Knuth_1996} to compute the new (perturbed) operating point\footnote{A two diode model can also be used \cite{Ishihara:IASTED:2011}} as summarized below.

\subsection{Outer-Loop Update using Lambert-W function}
\label{SSEC:lambertoverview}
In the simplest representation, the PV cell is assumed to be a superposition of the dark and illuminated current-voltage characteristics. Along with the series and shunt resistances, the I-V relationship of the PV cell is described here. In what follows, $I_{ph}$(A) denotes the photo-generated current, $I_0$(A) denotes the dark saturation current, $q$(Coulombs) denotes the electric charge carried by a single photon, $k$ denotes the Boltzmann constant ($J\cdot K^{-1}$), and $T$ denotes the cell temperature ($K$). Let us consider the following equations\footnote{For the sake of notational convenience, we omit the subscripts $k$ from $\beta_k$ and $k+1$ from $V_{k+1}$ and $I_{k+1}$ in Equation \ref{EQ:beta2}} to model the I-V characteristics of the PV cell:
\be
\label{EQ:beta}
V = \beta I
\ee
and
\be
\label{EQ:OneDiodeWithRsRsh}
I = I_{ph} - I_{01}\left(e^{\frac{(V+R_s I)}{V_{T_1}}}-1\right)-\frac{V+IR_s}{R_{sh}}
\ee
Substitution of (\ref{EQ:beta}) into (\ref{EQ:OneDiodeWithRsRsh}) yields: 
\be
\label{EQ:VvsBeta}
V = c_0^{-1}\beta \left(I_{ph}+I_{01}-I_{01}e^{\left(\frac{\beta+R_s}{\beta V_T}\right)} V\right)
\ee
where
\beas
c_0 &=& 1+ \frac{\beta +R_s}{R_{sh}}\\
\eeas
Equation (\ref{EQ:VvsBeta}) is of the form
\be
\label{EQ:mainPertEQ}
y = d_0 -d_1e^{\a_1 y}
\ee 
where
\beas
d_0 &=& c_0^{-1}\beta (I_{ph}+I_{01})\\
d_1 &=& c_0^{-1}\beta I_{01}\\
\a_1 &=& \frac{\beta+R_s}{\beta V_T}
\eeas
We can transform (\ref{EQ:mainPertEQ}) to $We^{W} = x$ where 
\beas
W &=& \a_1 d_0-\a_1y\\
x &=& \a_1d_1e^{\a_1d_0}
\eeas
Using the Lambert-W function \cite{Corless_Gonnet_Hare_Jeffrey_Knuth_1996}, we can obtain closed-form solutions with $y = V$ as follows: $y = \frac{1}{\a_1}\left(\a_1 d_0 -W(x)\right)$. The implication is that, given $\beta_k$ and the I-V parameters, the new (perturbed) operating point can be exactly computed. This enables one to rapidly simulate electrical performance under the time-varying non-uniform conditions. The presented approach above is scalable, robust, and readily extends to arbitrary circuit topologies. Given $N_p$ strings of $N_s$ cells connected in series per string, the single cell parameters $(I_{ph}, I_{01}, R_s, R_{sh}, V_T)$ scale into the corresponding module parameters $(N_pI_{ph}, N_pI_{01}, \frac{N_s}{N_p}R_s, \frac{N_s}{N_p}R_{sh}, N_sV_T)$.

\section{Modeling Solar Module and DC-DC Buck Boost Converter with Parasitic Losses}
\label{SEC:solarbuckboost}
\begin{figure}[t]
\begin{center}
\includegraphics[width=0.5\textwidth]{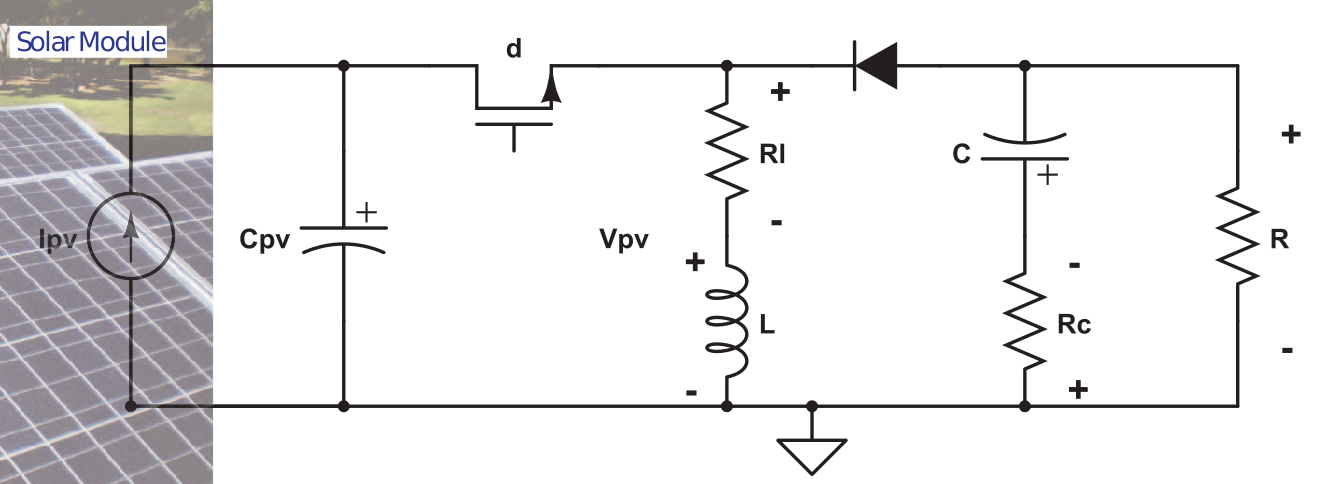} 
\end{center}
\caption{\enskip Solar Module and DC-DC Buck Boost Converter with Parasitic losses.}
\label{FIG:buck_boost_test-01}
\end{figure}

We consider a dynamical model of the PV-Buck-Boost system with parasitic resistances in the inductor, capacitor, and the \emph{on}-state of the MOSFET. The circuit diagram is shown in Fig. \ref{FIG:buck_boost_test-01}.

By representing the circuit elements as equations, we derive the dynamics of the system below.

\textbf{Switch-On Model:} When the switch is in the on-state, the equations become:
\bea
\label{EQ:Pos1}
\frac{d}{dt}\begin{bmatrix}v_{pv} \\ v_c \\ i_L\end{bmatrix} &=& \begin{bmatrix}0 & 0 & -\frac{1}{C_{pv}} \\ 0 & -\frac{1}{C(R+R_C)} & 0 \\ \frac{1}{L} & 0 & -\frac{(R_{on}+R_L)}{L}\end{bmatrix}\begin{bmatrix}v_{pv} \\ v_c \\ i_L\end{bmatrix} \nn \\ && + \begin{bmatrix} \frac{I_{pv}}{C_{pv}} \\ 0 \\ 0\end{bmatrix} \nn \\
&=& A_1x + h_1(x)
\eea
\textbf{Switch-Off Model:} When the switch is in the off-state, the equations become:
\bea
\label{EQ:Pos2}
\frac{d}{dt}\begin{bmatrix}v_{pv} \\ v_c \\ i_L\end{bmatrix} &=& \begin{bmatrix}0 & 0 & 0 \\ 0 & -\frac{1}{C(R+R_C)} & \frac{R}{C(R+R_c)} \\ 0 & -\frac{R}{L(R+R_C)} & -\frac{(R_L+R_d+R_c||R)}{L}\end{bmatrix}\begin{bmatrix}v_{pv} \\ v_c \\ i_L\end{bmatrix} \nn \\ && + \begin{bmatrix} \frac{I_{pv}}{C_{pv}} \\ 0 \\ -\frac{V_D}{L}\end{bmatrix} \nn \\ 
&=& A_2x + h_2(x)
\eea
\textbf{Averaged Model:} Using the averaged circuit model approach, the switched linear system can be approximated\footnote{The degree to which the nonlinear system approximates the switched linear system can be measured by application of the Baker-Campbell-Hausdorff formula.} by a single nonlinear system given by
\begin{equation}
\label{EQ:Avg_model}
\dot{x} = A(x,d) + dh_1 + (1-d)h_2
\end{equation}
\bes
A(x,d) \eqd dA_1+(1-d)A_2 = \begin{bmatrix}0 & 0 & -\frac{d}{C_{pv}} \\ 0 & -\frac{1}{C(R+R_C)} & \frac{(1-d)R}{C(R+R_C)} \\ \frac{d}{L} & -\frac{(1-d)R}{L(R+R_C)} & a_{33}\end{bmatrix}
\ees
and $a_{33} = -\frac{1}{L}\{d(R_{on}+R_L)+(1-d)(R_L+R_d+R_C\|R)\}$.  
In the above, $(v_{pv}, v_c, i_L)$ are the state variables, $d\in[0,1]$ is the control (duty-cycle) and the nonlinearities are due to the multiplicative control and state terms and the nonlinear function of the state variable: $I(v_{pv})$.

\section{Feedback Linearization Control}\label{SEC:FLC}
For controller design we consider (\ref{EQ:Avg_model}) but without parasitic resistances. However, in the simulation presented in Section \ref{SEC:results}, the controller is found to be robust even in the presence of parasitic resistances. The dynamics without the parasitic resistances are given by: 
\begin{equation}
\label{EQ:Three_state_dyn}
\begin{bmatrix}
\dot{v}_{pv} \\ \dot{v}_c \\ i_L \end{bmatrix} = \begin{bmatrix} 0 & 0 & -\frac{d}{C_{pv}} \\ 0 & -\frac{1}{CR} & \frac{1-d}{C} \\ \frac{d}{L} & -\frac{1-d}{L} & 0 \end{bmatrix} \begin{bmatrix} v_{pv} \\ v_c \\ i_L \end{bmatrix} + \begin{bmatrix} \frac{I_{pv}}{C_{pv}} \\ 0 \\ 0 \end{bmatrix}
\end{equation}
In the following, we discuss only the mechanics of the feedback linearization controller design.  We do not discuss the stability nor robustness properties. Readers interested in the theory should consult \cite{Su:1983}.  

\ni \textbf{Region I Controller: } Consider an operating point in region I as shown in Fig. \ref{FIG:IV}.  We assume a reference equivalent resistance is generated via (\ref{EQ:ReqDef2}) denoted by $R_{eq_{ref}}$.  In order to ensure tracking of $R_{eq_{ref}}$ we define the output variable: 
\bes
\yr(x) = R_{eq_{ref}} - R_{eq}(x)
\ees
Differentiating $\yr(x)$ with respect to time, we obtain
\begin{equation}
\label{EQ:Req_diff}
\yrd(x) = -\frac{d}{dt}\left(R_{eq}(x)\right) = -\frac{I_{pv}\dot{v}_{pv} - v_{pv}\dot{I}_{pv}}{I_{pv}^2}
\end{equation}
Application of the chain rule yields:
\begin{equation}
\label{EQ:Ipv_diff}
\dot{I}_{pv} = \frac{d}{dt}I_{pv}(v_{pv}) = \frac{\partial{I_{pv}}}{\partial{v_{pv}}}\dot{v}_{pv}
\end{equation}
Consider $N_s$ cells in series with identical parameters. If we have $N_p$ strings of $N_s$ cells each, in parallel, then
\begin{equation}
\label{EQ:solar_cell}
I_{pv} = I_{ph} - I_{01}(e^{\frac{v_{pv}+R_sI_{pv}}{V_T}}-1) - \frac{v_{pv}+R_sI_{pv}}{R_{sh}}
\end{equation}
where,
\begin{equation}
\begin{aligned}
\label{EQ:scaling_parms}
I_{ph} = N_pI_{ph}^{(c)} \\
I_{01} = N_pI_{01}^{(c)} \\
R_x = \frac{N_s}{N_p}R_s^{(c)} \\
R_{sh} = \frac{N_s}{N_p}R_{sh}^{(c)} \\
V_{T} = N_s V_{T}^{(c)}
\end{aligned}
\end{equation}
Thus we have,
\begin{align}
\begin{split}\label{eq:1}
    \frac{\partial{I_{pv}}}{\partial{v_{pv}}} ={}& -\frac{I_{01}}{V_T}e^{\frac{v_{pv}+R_sI_{pv}}{V_T}}(1+R_s\frac{\partial{I_{pv}}}{\partial{v_{pv}}}) \\ \\
         & - \frac{1}{R_{sh}}(1+R_s\frac{\partial{I_{pv}}}{\partial{v_{pv}}}) \\
\end{split}\\
\begin{split}
    \implies {}& \frac{\partial{I_{pv}}}{\partial{v_{pv}}}(1+\frac{I_{01}R_s}{V_T}e^{\frac{v_{pv}+R_sI_{pv}}{V_T}}+\frac{R_s}{R_{sh}}) \\ \\
         & = -\frac{I_{01}}{V_T}e^{\frac{v_{pv}+R_sI_{pv}}{V_T}} - \frac{1}{R_{sh}}
\end{split}
\end{align}

Hence we get,
\begin{equation}
\label{EQ:dI_dV_calc}
\scriptstyle
\frac{\partial{I_{pv}}}{\partial{v_{pv}}}(I_{pv},v_{pv})  = \\
-\frac{R_{sh}I_{01}e^{\frac{v_{pv}+R_sI_{pv}}{V_T}}+V_T}{V_TR_{sh}+I_{01}R_sR_{sh}e^{\frac{v_{pv}+R_sI_{pv}}{V_T}}+V_TR_s}
\end{equation}

We may now evaluate Equation \ref{EQ:Req_diff}
\begin{align}\label{EQ:ydot}
\dot{y} &= -\frac{1}{I_{pv}^2}(I_{pv}-v_{pv}\frac{\partial{I_{pv}}}{\partial{v_{pv}}})\dot{v}_{pv} \nonumber \\ \\
&= -\frac{1}{I_{pv}^2}(I_{pv}-v_{pv}\frac{\partial{I_{pv}}}{\partial{v_{pv}}})(-\frac{d}{C_{pv}}i_L+\frac{I_{pv}}{C_{pv}}) \nonumber \\ \\
&= -ky
\end{align}
Setting $\dot{y}=-ky$ where $k>0$ and solving for $d$, we have
\bea
\label{EQ:d_cal}
\ds
-\frac{d}{C_{pv}}i_L+\frac{I_{pv}}{C_{pv}} &=& \frac{kyI_{pv}^2}{I_{pv}-v_{pv}\frac{\partial{I_{pv}}}{\partial{v_{pv}}}} \notag \\
\frac{di_L}{C_{pv}} &=& \frac{I_{pv}}{C_{pv}} - \frac{kyI_{pv}^2}{I_{pv}-v_{pv}\frac{\partial{I_{pv}}}{\partial{v_{pv}}}} \notag \\
 &=& \frac{I_{pv}(I_{pv}-v_{pv}\frac{\partial{I_{pv}}}{\partial{v_{pv}}})-kyI_{pv}^2C_{pv}}{C_{pv}(I_{pv}-v_{pv}\frac{\partial{I_{pv}}}{\partial{v_{pv}}})}\frac{C_{pv}}{i_L} \nn \\
\implies d &=& \frac{I_{pv}(g(v_{pv})-kyI_{pv}C_{pv})}{i_Lg(v_{pv})}
\eea
where, $\ds g(v_{pv}) = I_{pv}-v_{pv}\frac{\partial{I_{pv}}}{\partial{v_{pv}}}$

\ni \textbf{Region II Controller: } Consider an operating point in region II as shown in Fig. \ref{FIG:IV}.  In this case, we assume a reference equivalent conductance is generated via (\ref{EQ:Geq}) denoted by $G_{eq_{ref}}$.  In order to ensure tracking of $G_{eq_{ref}}$ we define the output variable: 
\bes
\yg(x) = G_{eq_{ref}} - G_{eq}(x)
\ees
Proceeding as above, we may solve for the feedback linearizing control law that guarantees the output error converges exponentially to the origin.  

\ni \textbf{Complete Control Law: }We combine the results of region I and II into the following globally valid control law: 
\be
\label{EQ:dGlobal}
d_{flc} = \ds\left\{\ba{ll} \frac{1}{i_L} \left(I_{pv} - \frac{k \yr I_{pv}^2 C_{pv}}{g(v_{pv})}\right)& \mbox{ if Region I} \\
\frac{1}{i_L} \left(I_{pv} + \frac{k \yg v_{pv}^2 C_{pv}}{g(v_{pv})}\right) & \mbox{ if Region II}\ea \right.
\ee

\section{Simulation Results}
\label{SEC:results}
\begin{figure}[t]
\begin{center}
\includegraphics[width=0.5\textwidth]{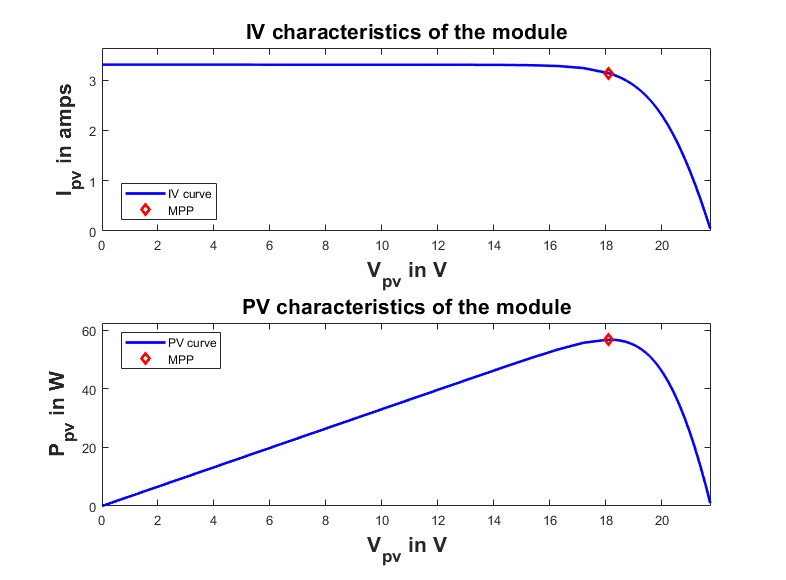}
\end{center}
\caption{\enskip I-V and P-V characteristics of the PV module considered.}
\label{FIG:pv_characteristics}
\end{figure}

\begin{figure}[t]
\begin{center}
\includegraphics[width=0.5\textwidth]{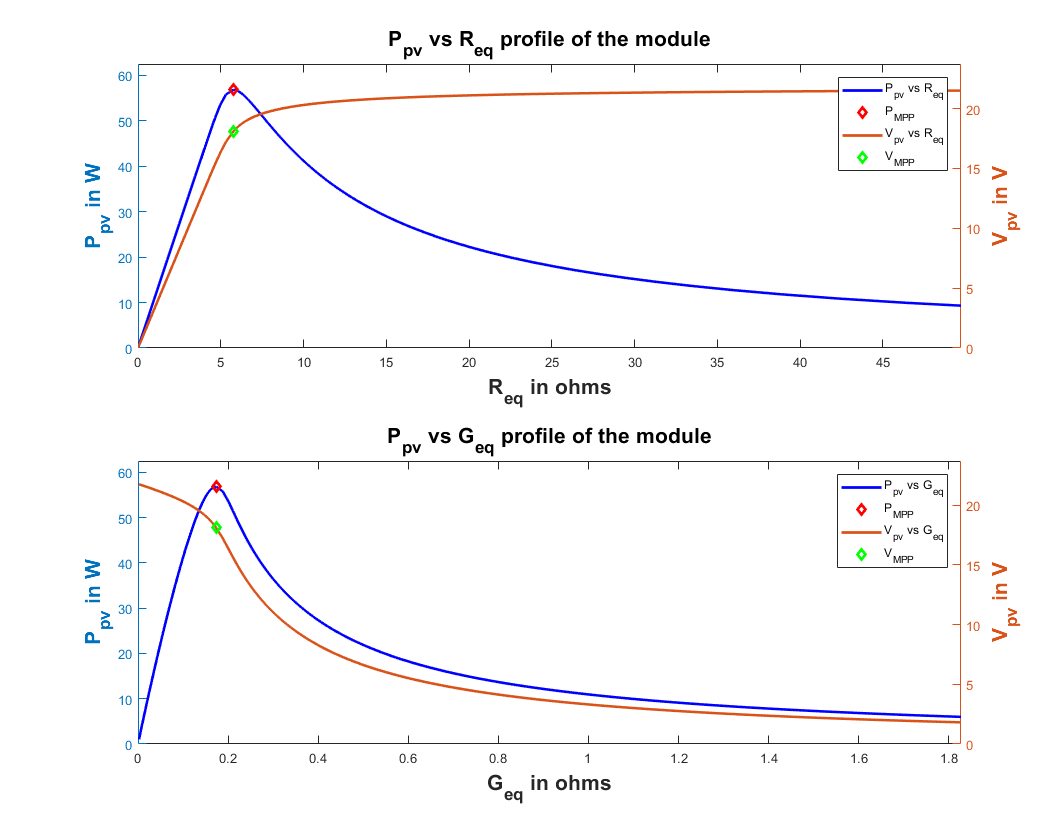}
\end{center}
\caption{\enskip $R_{eq}$ and $G_{eq}$ profiles of the PV module in relation to $P_{pv}$ and $V_{pv}$}
\label{FIG:req_geq}
\end{figure}

In this section we simulate the proposed feedback linearization-based control algorithm obtained in Equation \ref{EQ:dGlobal} on the PV-buckboost platform. The electrical parameters of a Kyocera PV module ($N_s$=36, $N_p$=1) employed are provided here: $R_s = 0.01$ (Ohms), $R_{sh} = 150$ (Ohms), $I_{01} = 1.9795\cdot 10^{-10}$ (A), and $I_{ph} = 3.31$ (A). For the DC-DC Buck-Boost converter, the following parameters are used: $C   = 220e^{-6} (F)$, $ L   = 3e-3 (H)$, $R   = 10 (\Omega)$, and $C_{pv} = 1e-3 (F)$.  We take the sampling period of the buck-boost converter to be $T_s =  1/100000 (s)$. Under the standard test conditions, the module characteristics are shown in Figure \ref{FIG:pv_characteristics}.
The MPP is achieved when the operating voltage is at $V_{mpp} \approx 18.1$ V. Therefore, region 1 can be understood to span the interval [0, $V_{mpp}$) and region 2 would span the interval [$V_{mpp}$, $V_{oc}$]. The equivalent resistance for region 1 (and conductance for region 2) can be seen from Figure \ref{FIG:req_geq}. The magnitude of $R_{eq}$ in region 1 and the magnitude of $G_{eq}$ in region 2 can be obtained from the portions of the graph that are left to the MPP. These magnitudes can provide cues to determine the step sizes associated with the iterative outer loop updates. Further, in order to adapt convergence based on the proximity to the MPP, the perturbation step sizes ($\Delta R_{eq}$ and $\Delta G_{eq}$) are chosen to be proportional to estimated slope $\frac{\partial P_{pv}}{\partial V_{pv}}$. In this work, the PV-buckboost model was simulated in MATLAB with the above parameters for different initial conditions.
The feedback linearization-based control converged to the correct Maximum Power Point by tracking the outer-loop set points iteratively without prior knowledge of the MPP. In order to test for robustness, we first note that the feedback controller derived in \ref{EQ:d_cal} does not account for parasitic resistances $(R_c, R_L, R_{on}, R_d)$. However, for this simulation the values of $(R_c, R_L, R_{on}, R_d)$ were set to $(1, 1, 1, 1000)$.
\begin{figure}[t]
\begin{center}
\includegraphics[width=0.5\textwidth]{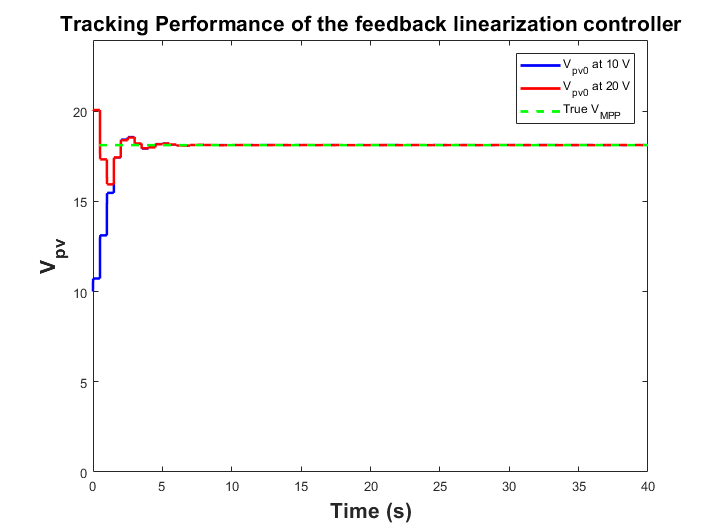}
\end{center}
\caption{\enskip Performance of Feedback Linearization Controller given in (\ref{EQ:dGlobal}) for two initial conditions.}
\label{FIG:FL_performanceTwoICs}
\end{figure}
The simulated trajectories starting from two initial conditions ($V_{pv}$ = 10 V, $V_{pv}$ = 20 V) are depicted in Figure \ref{FIG:FL_performanceTwoICs}. It can be noted that the system trajectory converges to the MPP even in the presence of parasitic effects, demonstrating the robustness of the controller. Further, the choice of step sizes proportional to the estimated slope of the PV curve ensures that the system trajectory does not oscillate about the MPP. 

\section{Conclusion}
\label{SEC:conclusion}
This work presented a novel method for MPPT by a combination of an equivalent resistance tracking mechanism achieved by feedback linearization of buck-boost converter dynamics. The analytical determination of the outer-loop set points using the Lambert-W function is discussed based on PV diode models. The buck-boost dynamics are derived for different switch positions and the average dynamics is formulated. A feedback linearization-based control law is derived to track the reference signal. The adaptive outer-loop set points and the exponential inner-loop tracking ensures rapid convergence to the maximum power point of the solar module. Simulation results are presented with different initial conditions. Results indicate that the proposed approach enables robust and stable tracking. Future work will investigate the comparison of existing techniques to the proposed approach under time-varying changes in plant parameters in simulation and experiment.

\printbibliography   
\end{document}